\documentclass[epj]{svjour}
\usepackage{graphicx}

\begin{document}

\author{V.P. Gusynin\thanks{On leave from Bogolyubov 
Institute for Theoretical Physics, 03143, Kiev, Ukraine.
\email{vgusynin@bitp.kiev.ua}}  
\and V.~A.~Miransky\thanks{On leave from Bogolyubov 
Institute for Theoretical Physics, 03143, Kiev, Ukraine.}
\fnmsep
\thanks{
\email{vmiransk@uwo.ca}}} 

\institute{Department of Applied Mathematics, University of
Western Ontario, London, Ontario N6A 5B7, Canada
}

\title{Thermal conductivity and competing orders
in d-wave superconductors}

\abstract{
We derive the expression for the thermal conductivity $\kappa$ in
the low-temperature limit $T \to 0$ in d-wave superconductors, taking 
into account the presence of competing orders such as spin-density wave,
$is$-pairing, etc.\,. The expression is used for analyzing recent 
experimental data in $\rm{La}_{2-{\it x}}\rm{Sr}_{\it x}\rm{CuO}_4$.
Our analysis strongly suggests that competing orders can be responsible 
for anomalies in behavior of thermal conductivity observed in those 
experiments.}


\maketitle

The existence of four nodal points in d-wave superconductors provides
rich and, sometimes, controllable
dynamics of quasiparticle excitations at zero temperature. In
particular, the expressions for electrical, thermal, and spin
conductivity simplify considerably in the universal-limit
$\omega \to 0, T \to 0$ \cite{univ,DL}. It is noticeable that the
role of the thermal conductivity $\kappa$ is special: 
while vertex and/or Fermi-liquid corrections modify the bare,
"universal", values of both electric and spin conductivities,
the universal value of the thermal conductivity is not influenced 
by them \cite{DL}. It is:
\begin{equation}
\frac{\kappa_0}{T}=\frac{k_B^2}{3}\frac{v_F^2+v_\Delta^2}{v_Fv_\Delta},
\label{kappa}
\end{equation}
where $v_F$ is a Fermi velocity, $v_\Delta$ is a gap velocity,
and $k_B$ is the Boltzmann constant (we use units with $\hbar = c =1$).  
The basis for such a remarkably simple expression is that there is
a finite density of states $N(0)$ of {\it gapless} quasiparticles
down to zero energy \cite{DL,Gorkov}:
\begin{equation}
N(0)=\frac{2}{\pi^2v_Fv_\Delta}\Gamma_0\ln\frac{p_0}{\Gamma_0},
\label{N}
\end{equation}
where $\Gamma_0 \equiv \Gamma(\omega\to 0)$, with $\Gamma(\omega)$
an impurity scattering rate, and 
$p_0 = \sqrt{\pi v_Fv_{\Delta}}/a$
is an ultraviolet momentum cutoff ($a$ is a lattice constant)
\cite{DL}. Note that expression (\ref{kappa}) itself is valid in the 
so-called
``dirty'' limit, $T\ll\Gamma_0$. Therefore,
although this expression does not contain 
$\Gamma_0$ explicitly, a nonzero $\Gamma_0$ is crucial both for
Eqs.(\ref{kappa}) and (\ref{N}). 

But what will happen if those quasiparticles become gapped? One 
may think that in that case both $N(0)$
and $\kappa_0$ are zero. However, as will be shown in this paper,
they both are finite even in that case, if the impurity
scattering rate is non-zero. In fact, it will be shown that
they are:
\begin{equation}
\frac{\kappa^{(m)}_0}{T}=\frac{k_B^2}{3}\frac{v_F^2+v_\Delta^2}{v_Fv_\Delta}
\frac{\Gamma_0^2}{\Gamma_0^2+m^2}
\label{kappa_m}
\end{equation}
and
\begin{equation}
N_m(0)=\frac{2}{\pi^2v_Fv_\Delta}\Gamma_0\ln\frac{p_0}
{\sqrt{\Gamma_0^2+m^2}},
\label{density_m}
\end{equation}
where $m$ a quasiparticle gap.
The noticeable point is that, for all values of the gap up to 
$m\simeq\Gamma_0$, the suppression of both thermal
conductivity and quasiparticle 
density is mild: $\kappa_0/\kappa^{(m)}_0$ and $N(0)/N_{m}(0)$ are of
order one. However,
the suppression in thermal conductivity
rapidly becomes strong as $m$ crosses this threshold. 
The second noticeable point is that, as we will discuss below, the gap
$m$ plays here a universal role and may represent different competing
orders in d-wave superconductors, such as as spin density wave,
charge density wave, $is$-pairing, etc.\,. Although their dynamics are
different, expressions (\ref{kappa_m}) and (\ref{density_m}) for
$\kappa^{(m)}_0$ and $N_m$ are the same. This happens
because, first, all those gaps $m$ correspond to different 
types of "masses"
in the Dirac equation describing nodal quasiparticle excitations, and,
secondly, unlike electric and spin conductivities,
the thermal conductivity $k^{(m)}_0$ is blind with respect to 
quantum numbers distinguishing those masses. 

The expression $k^{(m)}_0$
corresponds to the dirty limit when $T \ll \Gamma_0$. 
In d-wave superconductors, $\Gamma_0$ can be as large as of order
1 K or even 10 K, and $k^{(m)}_0$ can be an important measurable
characteristic there. Recently, two experimental groups
have observed an anomalous behavior in the thermal conductivity
in underdoped $\rm{La}_{2-{\it x}}\rm{Sr}_{\it x}\rm{CuO}_4$ 
(Refs. \cite{Ando1,Ando2} and Refs. \cite{Can1,Can2}).
One of the most interesting observations of experiment \cite{Can1}
is that at very low temperatures the value of the thermal conductivity
in underdoped $\rm{La}_{2-{\it x}}\rm{Sr}_{\it x}\rm{CuO}_4$ (LSCO) is 
{\it less} than the absolute minimum $k_{min}/T = 2k^{2}_B/3$ of
expression (\ref{kappa}) for $k_0/T$, corresponding
to the isotropic case with $v_F = v_{\Delta}$.
This puzzle can be naturally explained by utilizing the
modified expression (\ref{kappa_m}) with a nonzero $m$ describing
a competing order in the superconducting phase. We will discuss
this and other results of experiments \cite{Ando1,Ando2,Can1,Can2} below. 
 
At subkelvin temperatures relevant to the low-$T$ heat conduction
experiments, we will use the continuum, low-energy, description
for the nodal quasiparticles in the $d$-wave state. At each node,
the quasiparticles are described by a two-component Nambu field. 
It will be convenient, following Ref. 
\cite{Herbut}, to utilize four-component fields, by combining
Nambu fields corresponding to the nodes within each
of the two diagonal pairs. Thus we have two four-component
Dirac fields. The corresponding representation for three Dirac
matrices is 
\begin{equation}
\gamma_0=\sigma_1\otimes I,\quad \gamma_1=-i\sigma_2\otimes \sigma_3,
\quad \gamma_2=i\sigma_2\otimes\sigma_1,
\label{matrices}
\end{equation} 
where $\sigma_i$ are the Pauli matrices and while the first factor in the
tensor product acts in the subspace of the nodes in a diagonal pair, the
second factor acts on indices inside a Nambu field.
The matrices satisfy the algebra
$\{\gamma_\mu,\gamma_\nu\}=2g_{\mu\nu}$, $g_{\mu\nu}=
(1,-1,-1)$, $\mu,\nu=0,1,2$. 

We will consider quasiparticle gaps with 
the matrix structure $O_i=(I,i\gamma_5,
\gamma_3,\gamma_3\gamma_5)$. Here the matrices $\gamma_3$ and
$\gamma_5$, anticommuting with matrices $\gamma_\nu$, are
\begin{equation}
\gamma_3=i\sigma_2\otimes\sigma_2,\quad \gamma_5=\sigma_3\otimes I.
\end{equation} 
Then, for each of the two four-component Dirac fields, the bare Matsubara
Green's function can be written as
\begin{equation}
G_0(i\omega_n, {\vec k})=\frac{1}{i\omega_n\gamma_0-v_Fk_1\gamma_1-
v_\Delta k_2\gamma_2-m_iO_i}.
\end{equation}
Therefore, different gaps $m_i$ correspond to different types of
Dirac masses. As was pointed out in Refs. \cite{Sachd1,TVF,KP,Herbut},
these gaps represent different competing orders in low energy limit.
In particular, the mass $m_1$, with  
$O_1=I$, describes the (incommensurate) cos spin-density-wave\\ (SDW),
and the mass $m_2$, with $O_2=i\gamma_5$, describes sin SDW. The
masses $m_3$ and $m_4$, with $O_3=\gamma_3$ and 
$O_4=\gamma_3\gamma_5$, correspond to the $id_{xy}$-pairing and 
the $is$-pairing, respectively. One can also consider a gap corresponding
to the charge-density-wave (CDW). In that case, one should introduce
a Dirac mass term mixing the four-component Dirac fields
corresponding to the two different diagonal pairs of the nodes. For
simplicity, we will not consider it in this letter.    

The scattering on impurities can be 
taken into account by introducing a Matsubara self-energy $\Sigma
(i\omega_n)$, so that the dressed Green's function becomes 
$G(i\omega_n, {\vec k})=G_0(i\omega_n-\Sigma(i\omega_n), {\vec k})$.
As usual, retarded Green's function is obtained by analytically
continuing Green's function $G$,
$G^R(\omega, {\vec k})=G(i\omega_n\rightarrow\omega+i\epsilon, 
{\vec k})$, and the impurity scattering rate is defined as
$\Gamma(\omega)=
-{\rm Im\Sigma^R(\omega)}$. At low temperatures we take $\Gamma_0\equiv
\Gamma(\omega\to0)$. The size of $\Gamma_0$ depends on the impurity
density $n_{imp}$ as well as on the scattering phase shift $\delta$.
Solving the Schwinger-Dyson equation for the self-energy in the 
self-consistent $t$-matrix approximation, one can find 
that in the unitary limit ($\delta=\pi/2$)
the equation determining $\Gamma_0$  for a nonzero $m_i$
has the form \cite{FranzVafek}
\begin{eqnarray}
\Gamma_0^2=\pi^2v_Fv_\Delta\tilde{\Gamma}\left[N_f\ln\frac{p_0^2}
{\Gamma_0^2+m_i^2}\right]^{-1},
\end{eqnarray}
where $N_f$ is the number of four-component Dirac fields and
$\tilde{\Gamma}=n_{imp}/\pi\rho_0$ with $\rho_0$ the normal
state density of states. 
Since $v_\Delta
\sim\Delta_0$, the magnitude of the superconducting gap, 
the scattering rate $\Gamma_0$ is proportional to\\
$\sqrt{\Delta_0\tilde{\Gamma}} \sim 
\sqrt{\Delta_0 n_{imp}}$.

The longitudinal dc thermal conductivity is calculated by means of 
the Kubo formula. In the bubble approximation, following the standard 
procedure, it can be expressed through the quasiparticle spectral 
function $A(\omega,{\vec k})$ as follows
\begin{eqnarray}
\kappa^{(m)}&=&\frac{\pi N_f}{8k_BT^2}\int_{-\infty}^\infty
\frac{d\omega\omega^2}
{\cosh^2\frac{\omega}{2k_BT}}\int\frac{d^2k}{(2\pi)^2}\nonumber\\
&\times&\left\{v_F^2tr\left[\gamma_1 A(\omega,{\vec k})
\gamma_1A(\omega,{\vec k})\right]\right.\nonumber\\
&+&\left. v_\Delta^2tr\left[\gamma_2A(\omega,{\vec k})\gamma_2A
(\omega,{\vec k})\right]\right\}.
\label{kappa_general}
\end{eqnarray}
Here the spectral function is given by the discontinuity of the fermion
Green's function
\begin{equation}
A(\omega,{\vec k})=-\frac{1}{2\pi i}\left[G^R(\omega+i\epsilon, {\vec k})
-G^A(\omega-i\epsilon, {\vec k})\right].
\end{equation}
With Green's function at hand, we can calculate $A(\omega,{\vec k})$.
For example, for the gap proportional to the unit Dirac matrix, it has the
form ($m\equiv m_1$) \cite{FGI}
\begin{eqnarray}
A(\omega,{\vec k})&=&\frac{\Gamma_0}{2\pi E}\left[\frac{\gamma_0E-v_Fk_1
\gamma_1-v_\Delta k_2\gamma_2+m}{(\omega-E)^2+\Gamma_0^2}\right.\nonumber\\
&+&\left.\frac{\gamma_0E+v_Fk_1\gamma_1+v_\Delta k_2\gamma_2-m}
{(\omega+E)^2+\Gamma_0^2}\right],
\label{spec_density}
\end{eqnarray}
where $E({\vec k})=\sqrt{v_F^2k_1^2+v_\Delta^2k_2^2+m^2}$ is the 
quasiparticle energy. Substituting the last expression in
Eq.(\ref{kappa_general}) and taking the limit $T\to0$, we arrive at
\begin{eqnarray}
\frac{\kappa^{(m)}_0}{T}&=&\frac{2\pi N_fk_B^2}{3}\int\frac{d^2k}
{(2\pi)^2}\frac{\Gamma_0^2}{(E^2+\Gamma_0^2)^2}\nonumber\\
&=&\frac{k_B^2N_f}{6}\frac{v_F^2+v_\Delta^2}{v_Fv_\Delta}
\frac{\Gamma_0^2}{\Gamma_0^2+m^2},
\label{kappa0-nonuniv}
\end{eqnarray}
i.e., we derived expression (\ref{kappa_m}) for the thermal
conductivity (in which $N_f = 2$).
The result for three other gaps, $m_2$, $m_3$, and
$m_4$, introduced above, is the same.

With the spectral function (\ref{spec_density}),
the density of states (per spin)
\begin{equation}
N_m(\omega)=\frac{1}{2}\int\frac{d^2k}{(2\pi)^2}tr\left[\gamma_0A
(\omega,{\vec k})\right]
\end{equation}
is easily calculated 
\begin{eqnarray}
N_m(\omega)&=&\frac{N_f}{2\pi^2v_Fv_\Delta}\left[\Gamma_0\ln\frac{p_0}
{\sqrt{\Gamma_0^2+(\omega-m)^2}}\right.\nonumber\\
&+&\left.\Gamma_0\ln\frac{p_0}
{\sqrt{\Gamma_0^2+(\omega+m)^2}}\right.\nonumber\\
&+&\left.|\omega|\left(\frac{\pi}{2}+\tan^{-1}\frac{\omega^2-m^2-\Gamma_0^2}
{2|\omega|\Gamma_0}\right)\right].
\label{density_omega}
\end{eqnarray}
It yields expression (\ref{density_m}) for the density
of states with zero energy. Therefore, in the presence of impurities, the
quasiparticle band survives even for a finite $m$. 
\footnote{Note that in 
the absence of impurities [$\Gamma_0=0$], we would
get
$N_m(\omega)|_{\Gamma_0=0}=({N_f}/{2\pi v_Fv_\Delta})|\omega|
\theta(\omega^2-m^2)$,
i.e., in that case, the mass $m$ would lead to a gap in the density of
states.}
The physical reason for this is the formation of impurity bound states 
inside the gap \cite{Balatsky}. Overlap between these states leads to
impurity band supporting the quasiparticle heat (and electric) current.

The observation of a residual linear in $T$ term in the thermal
conductivity in cuprates ($\rm{YBa}_2\rm{Cu}_3\rm{O}_{7-\delta}$ (YBCO)
\cite{Taillefer}, $\rm{Bi}_2\rm{Sr}_2\rm{CaCu}_2\rm{O}_8$ (Bi-2212) 
\cite{Behnia} as well as LSCO \cite{Can1}) is usually interpreted
as a direct consequence of nodes in the gap. However, as it follows from 
Eq.(\ref{kappa0-nonuniv}), a subdominant order parameter, leading to
a gap for nodal quasiparticles, does not exclude such a linear term in the
thermal
conductivity, although the latter does not have a universal form anymore. 
\footnote{Although the fact that
opening of a gap for nodal quasiparticles leads to changes in
$\kappa_0^{(m)}$ is natural
from physical viewpoint, there has been a controversy concerning
this point in the literature.
For example, in the recent paper \cite{FranzVafek} the authors claim
that in the limit $T\to0$ the universal expression for the 
thermal conductivity, Eq. (\ref{kappa}), survives even for gapped
quasiparticles. Expression (\ref{kappa0-nonuniv}) derived
above clearly shows that this is not the case.}

Thus we conclude that nonperturbative dynamics, responsible for
the creation of competing orders in the supercritical phase,
can violate the universality in the thermal conductivity in the
low temperature limit $T \to 0$. 
Recent experiments indicate
that
the existence of such competing orders is quite possible 
\cite{exp}. Several theoretical models have been proposed to
describe this phenomenon (for a review,
see Ref. \cite{Sachd}). As we will now discuss, using the expression
for the thermal conductivity derived above, this phenomenon
can be relevant
for understanding recent experiments 
in $\rm{La}_{2-{\it x}}\rm{Sr}_{\it x}\rm{CuO}_4$ 
\cite{Ando1,Ando2,Can1,Can2}.

The measurements of the thermal conductivity in \\
LSCO at low temperature
\cite{Ando1,Ando2,Can1,Can2} showed the following 
characteristic features:

a) At subkelvin temperatures, the value of $\kappa/T$ decreases with
$x$ \cite{Ando1,Can1}. At temperature as low as 40 mK, the value of 
$\kappa/T$ in some underdoped samples is either less than
the absolute minimum 
$\kappa_{min}/T = 2k^{2}_B/3$ 
of expression (\ref{kappa}) 
(for $x = 0.06$) or quite close to it 
(for $x = 0.07$ and $x = 0.09$) \cite{Can1}. On the other hand, this
anomalous behavior in the thermal conductivity
disappears in overdoped samples ($x = 0.17$ and $x = 0.20$)
\cite{Can1}.

b) The evolution of $\kappa/T$ across optimum doping is smooth 
\cite{Ando1,Can1}.

c) The thermal conductivity is sensitive to magnetic field. While
in overdoped samples it increases with magnetic field, in underdoped
samples the thermal conductivity decreases with increasing magnetic
field \cite{Ando2,Can2}. The authors of Refs. \cite{Ando2,Can2} 
describe this as a field-induced thermal metal-to-insulator transition.

d) Although remaining smooth, the evolution of 
$\kappa/T$ across optimum doping becomes visibly faster with increasing 
magnetic field \cite{Ando2}.
    
The results of item a) can be easily understood if one assumes that
there exists a competing order, described by the Dirac mass $m$, in
the superconducting phase of underdoped LSCO. Then 
an appropriate value of $m$ in expression (\ref{kappa_m}) will provide
the necessary suppression of the thermal conductivity. The fact
that such an anomalous behavior in $\kappa/T$
disappears with increasing $x$, in overdoped samples,
can be understood if one assumes that the
dynamical
gap ("mass") $m$ decreases with increasing $x$. As to this assumption,
it is well known in quantum field theory that, indeed,
an increase of the fermion density often
suppresses a dynamical Dirac mass. The reasons for that are
simple. With increasing the fermion density, the screening
effects become stronger and the quasiparticle interactions become
weaker. In addition, at a sufficiently large quasiparticle density,
the energy gain from creating a gap $m$ in
the quasiparticle spectrum will be surpassed by the energy loss of
pushing up the energy of all states in the band above the gap.
In the case of the model
with Dirac fermions describing highly oriented pyrolytic graphite
(HOPG) \cite{Khv,GGMS}, this fact was explicitly shown in
Ref. \cite{GGMS}. Although the present system is quite different
from HOPG, that example supports plausibility of
this assumption.

It is tempting to speculate that the dynamical gap $m$ disappears
close to optimum doping ($x_0 = 0. 16$ in LSCO). A smooth evolution
of $\kappa/T$ across optimum doping then suggests that it could be 
a continuous phase transition with the scaling law of the form
$m \sim (x_c - x)^{\nu}$ in the scaling region with 
$0 < (x_c - x)/x_c \ll 1$, where the critical value $x_c \simeq x_0$. 
The critical index $\nu = 1/2$ would
correspond to the mean-field phase transition. In that case, there
would be a kink in expression (\ref{kappa_m}) at the critical
point $x=x_c$. Indeed, since the thermal conductivity 
(\ref{kappa_m}) depends on $m^2$, and there is a linear in $m^2$
term as $m^2 \to 0$, its derivative with respect to $x$ 
will have a finite discontinuity at $x = x_c$ for $\nu = 1/2$.
In the case of a non-mean-field continuum phase transition, with
$\nu > 1/2$, the evolution of $k/T$ across $x_c \simeq x_0$ would be 
smoother.

This picture, with appropriate modifications, can survive     
in the presence of a magnetic field. In particular, the fact that
in overdoped samples $\kappa$ increases with magnetic field as
$\sqrt{H}$ 
\cite{Ando2,Can2}, implies that the dynamics 
in a magnetic field in overdoped samples is apparently conventional.
Indeed, the $\sqrt{H}$ behavior is well described by
semiclassical models \cite{semicl}. This seems to suggest that there
is no gap $m$ (competing order) in overdoped samples.

The situation is different in underdoped samples. The magnetic
field enhances the suppression in $\kappa$ observed in the same
samples at zero field (item c) above). Moreover, the evolution
$\kappa/T$ across optimum doping becomes visibly faster with
increasing $H$ (item d)). This suggests that magnetic field
plays here the role of a catalyst, enhancing the gap $m$. For
sufficiently large values of $m$, the suppression in 
$\kappa$ will
be so large that a sample effectively becomes a thermal
insulator as was observed in experiments \cite{Ando2,Can2}. 

Microscopic dynamics responsible for creating
competing orders can be quite sophisticated \cite{Sachd}. This 
issue is outside the scope 
of this letter. Here we will only comment on the role of a magnetic field
as a catalyst in generating the gap $m$. In non-superconducting systems,
it is well known that a magnetic field is indeed a strong catalyst
in generating gaps (masses) for Dirac fermions \cite{GMS}. In
particular, this effect was studied in the model describing
HOPG \cite{Khv,GGMS}. It is clear, however, that the dynamics
in the vortex phase of d-wave superconductors is very
different and the question about the
relevance of a magnetic field for
generating (or enhancing) a quasiparticles gap there
is still open. For example, while the authors of
papers \cite{Laughlin,VMFT,Vish} 
believe that such a role for a magnetic field in that phase 
is plausible, the analysis of the authors of Ref. \cite{Li} 
indicates that the magnetic field can actually supress $id_{xy}$
and $is$ gaps in a $d$-wave state.

In this paper, we will use a heuristic approach and demonstrate that
the experimental data in Refs. \cite{Ando2,Can2} can be
qualitatively understood if one requires a gap
that is generated below a critical doping and increases
with a magnetic field. To make this point to be transparent,
we are looking for an ansatz for the gap $m(H,x)$
which would be as simple as possible. 
We assume that a) the phase transition at the critical doping
$x = 0.16$ is the mean-field (or nearly mean-field) one,
and b) the gap increases as $\sqrt{H}$ with the magnetic field
(such a scale covariant dependence of $m$ on $H$ was first considered in
$d$-wave superconductors in
Ref. \cite{Laughlin}). This leads us to the ansatz:
\begin{equation}
m(H,x)=\left({1-{x}/{0.16}}\right)^{1/2}\,\theta(0.16-x)(m_0+bE_H),
\label{ansatz}
\end{equation}
where $\theta$ is the step function, $E_H=\hbar v_F/2R\\
=(\hbar
v_F/2)\sqrt{eH/\hbar c}$
is a characteristic energy
scale in the presence of a magnetic field in the vortex state 
($2R$ is the average distance between vortices), and
$m_0$ and $b$ are free parameters. 
Taking $v_F=2.5\cdot 10^7 cm/s$ for LSCO cup\-ra\-tes
\cite{Can2,Can3},
we find $E_H=38K\cdot\sqrt{H(T)}$ where the field $H(T)$ is
taken in Teslas. The constant $b$ is of order $1$ (for
numerical calculations we take $b=2.2$). As to the parameter
$m_0$ that determines the gap for $H=0$,
it can be found from the ratio $\kappa/\kappa_0 =2/3$
(i.e., $\kappa/T\simeq 12\mu {\rm W K}^{-2}{\rm cm}^{-1}$)
for $x=0.06$, $H=0$ and $T \to 0$, as reported in Ref.\cite{Can1}.
Then, taking $\kappa/\kappa_0 = \kappa_{0}^{m}/\kappa_0$, with
$\kappa_{0}^{m}$ from equation (\ref{kappa_m}), we get
$m_0 = a\Gamma_0$ where the constant $a \simeq 0.9$.

Let us now calculate the thermal conductivity by using 
ansatz (\ref{ansatz}) for $m(H,x)$.
The impurity bandwidth for LSCO is estimated to be $\Gamma_0\simeq
25 K - 30 K $ \cite{Can2,Can3} which is two orders of magnitude larger
than for
very clean\\
$\rm{YBa}_2\rm{Cu}_3\rm{O}_{6.99}$ samples. While in clean 
$\rm{YBa}_2\rm{Cu}_3\rm{O}_{6.99}$ the scattering of quasiparticles
from vortices must be
taken into account, one can neglect 
the dependence of the width $\Gamma_0$ on the magnetic field (at least for
not very high fields) in the case of rather dirty LSCO. 
On the other hand, the presence of a circulating supercurrent around 
vortices in 
the vortex state can be taken into account in the semiclassical approach 
by making the Doppler shift in quasiparticle energies, $\omega\to\omega-
{\bf v}_s({\bf r}){\bf k}$, \cite{Volovik} (${\bf v}_s({\bf r})$ is the
superfluid velocity at a position ${\bf r}$ which depends on the form
of vortices distribution). In this case, the local thermal conductivity
$\kappa(r)$ has to be averaged over the unit cell of the vortex lattice
\cite{vortex-averaging}, 
\begin{equation}
\kappa(H,T)=\frac{1}{A}\int d^2r\,\kappa(r)=\int d\epsilon\,
{\cal P}(\epsilon)\kappa(\epsilon,T), 
\end{equation}
where 
\begin{equation}
{\cal P}(\epsilon)=\frac{1}{A}\int d^2r\delta(\epsilon-{\bf v}_s({\bf r})
{\bf k})
\end{equation}
 is the vortex 
distribution, and $A=\pi R^2$ is the area of the vortex unit cell.
We use the Gaussian distribution function ${\cal P}(\epsilon)=(1/
\sqrt{\pi}E_H)\exp[-\epsilon^2/E_H^2]$ which is believed to be the most 
suitable distribution in the presence of high disorder \cite{Leggett}. 
Thus we need to calculate
\begin{eqnarray}
\kappa(H,T)&=&\frac{\pi N_f}{8k_BT^2}\int_{-\infty}^\infty
\frac{d\omega\omega^2}{\cosh^2\frac{\omega}{2k_BT}}\int_{-\infty}^\infty 
d\epsilon{\cal P}(\epsilon)
\int\frac{d^2k}{(2\pi)^2}\nonumber\\
&\times&\left\{v_F^2tr\left[\gamma_1 A(\omega-\epsilon,{\vec k})
\gamma_1A(\omega-\epsilon,{\vec k})\right]\right.\nonumber\\
&+&\left. v_\Delta^2tr\left[\gamma_2A(\omega-\epsilon,{\vec k})\gamma_2A
(\omega-\epsilon,{\vec k})\right]\right\}
\label{kappa_averaged}
\end{eqnarray}
(compare with Eq.(\ref{kappa_general})).

Taking the limit $T\to0$ in the last equation, we arrive at the
following expression:
\begin{eqnarray}
\frac{\kappa(H,0)}{\kappa_0}&=&\frac{1}{2}\int_{-\infty}^\infty d\epsilon
{\cal P}(\epsilon) \left[1+\frac{\epsilon^2-m^2+\Gamma_0^2}{2|\epsilon|
\Gamma_0}\right.\nonumber\\
&\times&\left.\left(\frac{\pi}{2}-\tan^{-1}\frac{\Gamma_0^2+m^2-\epsilon^2}
{2|\epsilon|\Gamma_0}\right)\right],
\end{eqnarray}
where we normalized the thermal conductivity on the universal value
Eq.(\ref{kappa}).

\begin{figure}[h]
\centering{\includegraphics[width=8cm]{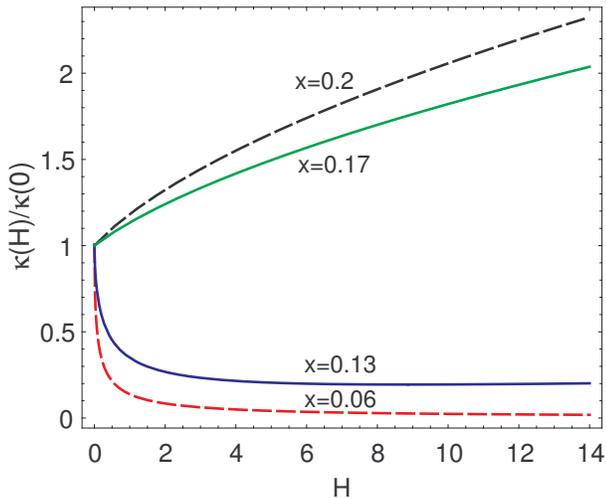}}
\caption{$\kappa(H)/T$ [normalized to the value $\kappa(0)/T$]
versus $H$ at $T=0$ and for the doping with the values 
$x=0.06, 0.13, 0.17, 0.2$. The 
impurity width is $\Gamma_0=25K$ (for the upper curve ($x=0.2$) 
$\Gamma_0=30K$).} 
\label{fig:1}
\end{figure}

\begin{figure}[h]
\centering{\includegraphics[width=8cm]{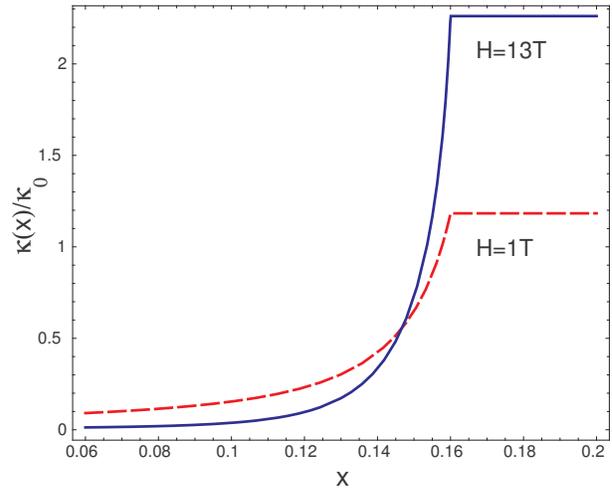}}
\caption{Doping dependence of the $T=0$ thermal conductivity $\kappa$
[normalized to the universal value $\kappa_0$]
for two values of the magnetic field $H=1T$ (dotted curve) and $H=13T$
(solid curve) and $\Gamma_0=25K$.
} 
\label{fig:2}
\end{figure}

In Figs.\ref{fig:1},\ref{fig:2} we present the ratio 
$\kappa(H,0)/\kappa_0$ calculated as
a function of the magnetic field $H$ (Fig.\ref{fig:1}) and the doping $x$  
(Fig.\ref{fig:2}). 
The form of these dependences is quite similar to that
of experimental data presented in Fig.2  of Ref. \cite{Can2} and in Fig.4 
of Ref. \cite{Ando2}, respectively.

At small values of the
doping, $x=0.06$ and $x=0.13$ (low curves in Fig.\ref{fig:1}), 
the thermal 
conductivity decreases with increasing field as a result of increasing the
gap $m(H)$. For supercritical values of the doping ($x=0.17; 0.2$ - upper
curves in Fig.\ref{fig:1}) the field dependence is approximately $\sqrt{H}$. 
This behavior is in accordance with the increase in quasiparticle
population due to the Volovik effect that is valid even for gapped 
quasiparticles \cite{BalatskyMao} when the vortex scattering is neglected.

Fig.\ref{fig:2} shows the dependence of $\kappa$ on the doping 
for two different values of the magnetic field. One can see the
suppression of $\kappa$ in the 
underdoped regime as a result of the presence of the
magnetic-field-induced gap. Note that both curves grow fast 
near the critical doping $x_c=0.16$ 
where the gap disappears. It is also noticeable that this growth is much
faster for the $H=13$ $ T$ curve
than that for the $H=1$ $T$ curve. These facts agree with the
experimental data \cite{Ando2} discussed in item d) above. 

Although the present analysis is based on the particular
ansatz (\ref{ansatz}) for $m(H,x)$, one can expect that
the main characteristics in the behavior of the thermal conductivity will
retain
qualitatively the same for a wide class of gaps $m(H,x)$
sharing the
features that they   
are generated below a critical doping and
increase with a magnetic field.

In conclusion, we derived the expression for the
thermal conductivity in d-wave superconductors in the presence of
competing orders.
The derived expression (\ref{kappa_m}) for
$\kappa^{(m)}_0/T$ is simple and transparent. We also analyzed
the dependence of the thermal conductivity on a magnetic field
and a doping in the vortex state. Our results strongly suggest
that the presence of competing orders can be crucial for
understanding recent experiments in LSCO
\cite{Ando1,Ando2,Can1,Can2}.
\begin{acknowledgement}
We thank E.V. Gorbar, D.V. Khveshchenko, V.M. Loktev, Yu.G. Pogorelov,
S.G. Sharapov and I.A. Shovkovy
for useful discussions. 
This work was supported by the Natural Sciences and
Engineering Research Council of Canada.  The
research of V.P.G. was also supported in part by the
SCOPES-projects 7~IP~062607 and 7~UKPJ062150.00/1 of Swiss NSF.
\end{acknowledgement}

\end{document}